\documentclass[showpacs,pre,twocolumn,floatfix,eqsecnum]{revtex4}
\usepackage{amsmath,epsfig}

\begin{document}

\title{Quantum theory of electromechanical noise and momentum transfer statistics}

\author{M. Kindermann}
\author{C.W.J. Beenakker}
\affiliation{Instituut-Lorentz, Universiteit Leiden, P.O. Box 9506, 2300 RA Leiden, The Netherlands}
\date{June 2002}
\begin{abstract}
A quantum mechanical theory is developed for the statistics of momentum transferred to the lattice by conduction electrons. Results for the electromechanical noise power in the semiclassical diffusive transport regime agree with a recent theory based on the Boltzmann-Langevin equation. All moments of the transferred momentum are calculated for a single-channel conductor with a localized scatterer, and compared with the known statistics of transmitted charge.
\end{abstract}
\maketitle

\section{Introduction}

Electrical current is the transfer of charge from one end of the conductor to the other. The statistics of this charge transfer was investigated by Levitov and Lesovik \cite{Lev93}. It is binomial for a single-channel conductor at zero temperature and double-Poissonian at finite temperature in the tunneling regime \cite{Lev01}. The second cumulant, the  noise power, has been measured in a variety of systems \cite{Bla00}. Ways of measuring the third cumulant have been proposed \cite{Gef02,Lev01}, but not yet carried out. 

Electrical current also transfers momentum to the lattice. The second cumulant, the electromechanical noise power, determines the mean square displacement of an oscillator through which a current is driven. It has been studied theoretically \cite{Yur90,Pre92,Sch95,Shy02} and is expected to lie within the range of sensitivity of nanomechanical oscillators \cite{Rou01}. No theory exists for higher order cumulants of the transferred momentum (which would determine higher cumulants of the oscillator displacement). It is the purpose of the present paper to provide such a theory. 

In the context of charge transfer statistics there exist two approaches: a fully quantum mechanical approach using Keldysh Green functions \cite{Lev93,Naz99} and a semiclassical approach using the Boltzmann-Langevin equation \cite{Nag02}. Here we take the former approach, to arrive at a quantum theory of momentum transfer statistics. As a test, we show that the second moment calculated from Keldysh Green functions coincides in the semiclassical limit with the result obtained from the Boltzmann-Langevin equation by Shytov, Levitov, and one of the authors \cite{Shy02}.

A calculation of the complete cumulant generating function of transferred momentum (or, equivalently, of oscillator displacement) is presented for the case of a single-channel conductor with a localized scatterer. The generating function in this case can be written entirely in terms of the transmission probability $\Gamma$ of the scatterer. In the more general multi-channel case one also needs knowledge of the wavefunctions. This is an essential difference from the charge transfer problem, which can be solved in terms of transmission eigenvalues for any number of channels.  At zero temperature the momentum statistics is binomial, just as for the charge. At finite temperature it is multinomial,  even in the limit $\Gamma \to 0$,  different from the double-Poissonian distribution of charge.

The outline of the paper is as follows. In Sec.\ \ref{sc:Formulation} we formulate the problem in a way that is suitable for further analysis. The key technical step in that section is a unitary transformation which  eliminates the dependence of the electron-phonon coupling Hamiltonian on the  (unknown) scattering potential of the disordered  lattice. The resulting coupling  Hamiltonian contains   the electron momentum  flow and the phonon displacement. In the next section we use that Hamiltonian to derive a general formula for the generating function of the  distribution of momentum transferred to a phonon  (as well as the distribution of phonon displacements). It is the analogue of the Levitov-Lesovik formula for the charge transfer distribution \cite{Lev93}. For a localized scatterer we can evaluate this statistics in terms of the scattering matrix. We show how to do this in Sec.\ \ref{sc:constantprofile}, and give an application to a single-channel conductor in Sec.\ \ref{sc:single}.  In Secs.\ \ref{sc:Keldysh} and \ref{sc:diffusive} we turn to the case that the scattering region extends throughout the conductor. We follow the Keldysh approach to derive a general formula for the generating function and  check its validity by rederiving the result of Ref.\ \cite{Shy02}. We conclude in Sec.\ \ref{sc:conclusion} with an order of magnitude estimate of higher order cumulants of the momentum transfer statistics.

\section{Formulation of the problem} \label{sc:Formulation}
The excitation of a phonon mode by conduction electrons is described by the
Hamiltonian
\begin{equation}
H=\Omega a^{\dagger}a+\sum_{i}{\bf p}_{i}^{2}/2m+\sum_{i}V[{\bf
r}_{i}-Q{\bf u}({\bf r}_{i})],\label{Hdef}
\end{equation}
where we have set $\hbar=1$. The phonon mode has annihilation operator $a$,
frequency $\Omega$, mass $M$, and displacement $ Q {\bf u}({\bf r})$, where $Q=(2 M\Omega)^{-1/2}(a+a^{\dagger})$ is the amplitude operator. The electrons have
position ${\bf r}_{i}$, momentum ${\bf p}_{i}=-i\partial/\partial{\bf
r}_{i}$, and  mass $m$. Electrons and phonons are coupled through the
ion  potential $V({\bf r})$. 
We assume zero magnetic field. 
Electron-electron interactions and the interactions of electrons and phonons
with an external electric field have also been omitted.

We assume that electrons and phonons are uncoupled at time zero and measure
moments of the observable $A$ of the phonons after they have been coupled to
the electrons for a time $t$. The operator $A(a,a^{\dagger})$ could be the
amplitude $Q$ of the phonon mode, its momentum
$P=-i (M  \Omega/2)^{1/2} (a-a^{\dagger})$, or its energy $\Omega a^{\dagger}a$.  The moment generating function for $A$  is
\begin{equation}
{\cal F}(\xi)=\sum_{m=0}^{\infty}\frac{\xi^{m}}{m!}\langle A^{m}(t)\rangle={\rm
Tr}\,e^{\xi A}e^{-iHt}\rho e^{iHt}.\label{eq:Fxidef}
\end{equation}
The initial density matrix $\rho=\rho_{e}\rho_{p}$ is assumed to factorize into
an electron and a phonon part.

We assume small displacements, so an obvious way to proceed would be to
linearize $V({\bf r}-Q{\bf u})$ with respect to the phonon amplitude $Q$. Such a procedure is
complicated by the fact that the resulting coupling $-Q{\bf u}\cdot\nabla V$ of
electrons and phonons depends on the ion  potential $V$. Because of momentum
conservation, it should be possible to find the momentum transferred by the
electrons to the lattice without having to consider explicitly the force
$-\nabla V$. In the semiclassical calculation of Ref.\ \cite{Shy02} that goal is
achieved by the continuity equation for the flow of electron momentum. The
unitary transformation that we now discuss achieves the same purpose in a
fully quantum mechanical framework.

What we need is a unitary operator $U$ such that
\begin{equation}
U^{\dagger}V[{\bf r}-Q{\bf u}({\bf r})]U=V({\bf r}).\label{eq:UVUdagger}
\end{equation}
For constant  ${\bf u}$  we have simply $U=\exp[-iQ{\bf u} \cdot{\bf p}]$. More generally, for space-dependent ${\bf u}$, we need to specify
the operator ordering (denoted by colons $:\cdots :$) that all position
operators ${\bf r}$ stand to the left of the momentum operators ${\bf p}$. We
also need to include a Jacobian determinant $||J||$ to ensure unitarity of $U$.
As shown in  App. \ref{app:U}, the  desired operator is
\begin{equation}
U=||J||^{1/2}:e^{-iQ{\bf u}({\bf r})\cdot{\bf
p}}:\,,\;\;J_{\alpha\beta}=\delta_{\alpha\beta}-Q \partial_{\alpha} u_{\beta}({\bf r}),\label{Uresult}
\end{equation}
 with
$\partial_{\alpha}\equiv\partial/\partial r_{\alpha}$. All this was for a
single electronic degree of freedom. The corresponding operator for many electrons is
$U=\prod_{i}U_{i}$, where $U_{i}$ is given by Eq.\ (\ref{Uresult}) with ${\bf
r}$, ${\bf p}$ replaced by ${\bf r}_{i}$, ${\bf p}_{i}$.

The Hamiltonian (\ref{Hdef}) transforms as $U^{\dagger}HU=H_{0}+H_{\rm int}$,
with
\begin{subequations} \label{eq:twoeq}
\begin{eqnarray} \label{H0def}
&&H_{0}=\Omega a^{\dagger}a+\sum_{i}[{\bf p}_{i}^{2}/2m+V({\bf
r}_{i})] , \\
&& H_{\rm int}= - Q F - \frac{1}{M} P \Pi +{\cal O}({\bf u}^2).\label{Hintdef}
\end{eqnarray}
\end{subequations}%
Here $F$ is the driving force of the phonon mode, 
\begin{equation} \label{eq:force}
F= \frac{1}{4m}\sum_{i}[u_{\alpha\beta}({\bf
r}_{i})p_{i\alpha}p_{i\beta}+p_{i\alpha}u_{\alpha\beta}({\bf
r}_{i})p_{i\beta}]+{\rm H.c},
\end{equation}
and $\Pi$ is the total electron momentum,
\begin{equation} \label{eq:Pi}
\Pi = {\textstyle{\frac{1}{2}}}\sum_{i}{\bf u}({\bf r}_{i})\cdot{\bf p}_{i}+{\rm
H.c.},
\end{equation}
weighted with the (dimensionless) mode profile ${\bf u}({\bf r})$. We have defined  the shear tensor $u_{\alpha\beta}=
\frac{1}{2}(\partial_{\alpha}u_{\beta}+\partial_{\beta}u_{\alpha})$.
The abbreviation H.c.\ indicates the Hermitian conjugate and a summation over
repeated cartesian indices $\alpha,\beta$ is implied. 

The interaction
Hamiltonian $H_{\rm int}$ is now independent of the ion  potential, as
desired. In the first term $-Q F$ we recognize the momentum flux tensor,
while the second term $-P \Pi/M$ is an inertial contribution to the momentum
transfer. The inertial contribution is of relative order $\Omega\lambda/v_{F}$ ($\lambda $ being the wavelength of the phonon and $v_F$ the Fermi velocity of the electrons)
and typically $\ll 1$.  In what follows we will neglect it.
We also neglect the terms in $H_{\rm int}$ of second and higher order in  ${\bf u}$, which contribute to order $\lambda_F/L$ to the generating function (with $L$ the length scale on which ${\bf u}$ varies). These higher order interaction terms account for  the momentum uncertainty of an electron upon a position measurement by the phonon.

If we apply the unitary transformation $U$ to the generating function
(\ref{eq:Fxidef}), we need to transform not only $H$ but also $A\rightarrow
U^{\dagger}AU=\tilde{A}$ and $\rho\rightarrow U^{\dagger}\rho
U=\tilde{\rho}$, resulting in
\begin{equation}
{\cal F}(\xi)={\rm Tr}\,e^{\xi\tilde{A}}e^{-it(H_{0}+H_{\rm int})}\tilde{\rho}
e^{it(H_{0}+H_{\rm int})}.\label{FHint}
\end{equation}
In App.\ \ref{app:U} we show that, quite generally, the distinction between $\rho,A$ and
$\tilde{\rho},\tilde{A}$ is irrelevant in the limit of a long detection time
$t$, and we will therefore ignore this distinction in what follows.

If ${\bf u}$ is smooth on the scale of $\lambda_F$, so that gradients of $u_{\alpha \beta}$ can be neglected, one can apply the effective mass approximation to the Hamiltonian  (\ref{eq:twoeq}). The ion potential $V=V_{\rm lat}+V_{\rm imp}$ is decomposed  into  a contribution $V_{\rm lat}$ from the periodic lattice and a contribution $V_{\rm imp}$ from impurities and boundaries that break the periodicity. The effects of $V_{\rm lat}$ can be incorporated  in an effective mass $m^*$ (assumed to be deformation independent \cite{Kon84,Fik78}) and a corresponding quasimomentum $p^*$.  The unperturbed Hamiltonian takes the usual  form
\begin{eqnarray}
H_{0}&=&\Omega a^{\dagger}a+\sum_{i}[{\bf p}^{\ast 2}_{i}/2m^*+V_{\rm imp}({\bf
r}_{i})].
\end{eqnarray}
  As shown in App.\ \ref{app:effmass},
the force operator in $H_{\rm int}$  is then expressed through the flow of quasi-momentum,
\begin{equation}
F= \frac{1}{m^*}\sum_{i} p^*_{i\alpha}u_{\alpha\beta}({\bf
r}_{i})p^*_{i\beta},
\end{equation}
whereas the inertial contribution is still given by Eq.\ (\ref{eq:Pi}) in terms of  the true electron momentum. 

\section{Momentum transfer statistics}  \label{sc:counting}
\subsection{Generating function}

A massive phonon mode absorbs  the momentum that electrons transfer to it without changing its displacement.  We may therefore define a statistics of momentum transfer to the phonons without back action on the electrons   by choosing  
 the observable $A=P=-i(M \Omega/2)^{1/2}(a-a^{\dagger})$ in Eq.\ (\ref{eq:Fxidef}) and taking the limit $M \to \infty$, $\Omega\to 0$ at fixed $M\Omega$. 
We  assume that the phonon mode is initially in the ground  state, so that $a \rho_p=0$.

We transform to the interaction picture by means of the identity
\begin{equation}
e^{iH_{0}t}e^{-iHt}={\cal T}\exp\left[-i\int_{0}^{t}dt'\,H_{\rm int}(t')\right], \label{interaction}
\end{equation}
where ${\cal T}$ denotes time ordering (earlier times to the right of later times) of the time dependent operator $H_{\rm int}(t)=e^{iH_{0}t}H_{\rm int}e^{-iH_{0}t}$. In the massive phonon limit we have   $H_{\rm int}(t)=-QF(t)$ with time independent $Q$ (since $Q$ commutes with  $H_0$ when $\Omega \to 0$). Eq.\ (\ref{eq:Fxidef}) takes the form
\begin{equation}
{\cal F}(\xi)=\langle {\cal T}_{\pm}\exp[-iQK_{-}(t)]e^{\xi P}\exp[iQK_{+}(t)]\rangle,\label{FHint2}
\end{equation}
where $K_{\pm}(t)=\int_{0}^{t}dt_{\pm}\,F(t_{\pm})$ and ${\cal T}_{\pm}$ denotes the Keldysh time ordering: times $t_{-}$ to the left of times $t_{+}$, earlier $t_{-}$ to the left of later $t_{-}$, earlier $t_{+}$ to the right of later $t_{+}$.

Taking the expectation value of the phonon degree of freedom  we find
\begin{eqnarray}
&& {\cal F}(\xi)=e^{\xi^2 M \Omega/2} \nonumber  \\
&&  \mbox{} \times\left\langle{\cal T}_{\pm}\exp\left[{\textstyle{\frac{1}{2}}}\xi(K_{-}+K_{+}) - \frac{ (K_+-K_-)^2}{4M \Omega} \right]\right\rangle. \;\;\;\;\;\;\;\;\label{Fxiint}
\end{eqnarray}
The factor $\exp(\xi^2 M\Omega/2)$ originates from the uncertainty $ (M \Omega)^{1/2}$ of the  momentum  of the phonon mode in the ground state (vacuum fluctuations). It is a time-independent additive contribution to the second cumulant and we   can omit it for long detection times. The quadratic term $\propto K_{\pm}^{2}/M \Omega$ becomes small for a small uncertainty $  (M\Omega)^{-1/2}$ of the  displacement in the ground state. It  describes a back action  of the phonon mode on the electrons that persists in the massive phonon limit. (A similar effect is known in the context of charge counting statistics \cite{Kin01}.) This term may be of importance in some situations, but we will not consider it here, assuming that the electron dynamics is insensitive to the vacuum fluctuations of the phonon mode. 

With these simplifications we arrive at a formula for the momentum transfer statistics,
\begin{equation}
{\cal F}(\xi)=\langle{\cal T}_{\pm}\exp[{\textstyle{\frac{1}{2}}}\xi K_{-}(t)] \exp[{\textstyle{\frac{1}{2}}}\xi K_+(t) ]\rangle,       \label{eq:Fxisimple}
\end{equation}
that is of the same form as the formula for charge counting statistics due to  Levitov and Lesovik \cite{Lev93},
\begin{equation}
{\cal F}_{\rm charge}(\xi)=\langle{\cal T}_{\pm}\exp[{\textstyle{\frac{1}{2}}}\xi J_{-}(t)] \exp[{\textstyle{\frac{1}{2}}}\xi J_+(t) ]\rangle.  \label{Fxicurr}
\end{equation}
The role of the integrated current $J(t)= \int_0^t{dt'\, I(t')}$ is taken in our problem  by the integrated force $K(t)$.

\subsection{Relation to displacement statistics}

Cumulants $\langle \langle \triangle P(t) \rangle \rangle$ of the momentum transferred in a time $t$ are obtained from the cumulant generating function $\ln {\cal F}(\xi) = \sum_n \langle \langle \triangle P(t)^n \rangle \rangle \xi^n/n!$. Cumulants $\langle \langle F(\omega)^n \rangle \rangle$ of the Fourier transformed force $F(\omega)= \int{dt\,e^{i\omega t} F(t)}$ then follow from the relation $\triangle P(t) = \int_0^t{dt' \, F(t')}$. The limit $t \to \infty$ of a long detection time corresponds to the low frequency limit,
\begin{equation}
\left\langle \left\langle \prod_{i=1}^n F(\omega_i) \right\rangle \right\rangle \to 2 \pi \delta\left(\sum_{i=1}^n \omega_i\right) \lim_{t \to \infty} \frac{1}{t} \left\langle \left\langle \triangle P(t)^n \right\rangle \right\rangle. 
\end{equation}

Cumulants of the Fourier transformed displacement $Q(\omega)$ of the oscillator follow from the phenomenological equation of motion 
\begin{equation}
Q(\omega) = R(\omega) F(\omega),\;R(\omega)= \frac{1}{\cal M}(\omega_c^2 - \omega^2 - i \omega \omega_c/{\cal Q})^{-1}.
\end{equation}
Here ${\cal M}$ is the motional mass, $\omega_c$ the characteristic frequency, and ${\cal Q}$ the quality factor of the oscillator. Since the force noise is white until frequencies that are typically $\gg \omega_c$, one has in good approximation
\begin{eqnarray} \label{eq:Qcum}
\left\langle \left\langle \prod_{i=1}^n Q(\omega_i) \right\rangle \right\rangle& =& 2 \pi \delta\left(\sum_{i=1}^n \omega_i\right) \prod_{j=1}^n R(\omega_j) \nonumber \\
&&  \mbox{} \times  \lim_{t \to \infty} \frac{1}{t} \left\langle \left\langle \triangle P(t)^n \right\rangle \right\rangle.
\end{eqnarray}

Optical or magnetomotive  detection of the vibration, as in Refs.\ \cite{Mey88,Tre96,Rou00}, measures the probability distribution $P(Q)$ of the displacement at any given time. The cumulants of $P(Q)$ are obtained by Fourier transformation of Eq.\ (\ref{eq:Qcum}), 
\begin{eqnarray} \label{eq:Qtom}
 &&\langle \langle Q^n \rangle \rangle = {\cal R}_n \lim_{t \to \infty} \frac{1}{t} \langle \langle \triangle P(t)^n \rangle \rangle,  \\
&& {\cal R}_n = \int{\frac{d\omega_1}{2\pi} \cdots \int{\frac{d\omega_n}{2\pi} R(\omega_1) \cdots R(\omega_n) 2\pi \delta\left(\sum_{i=1}^n \omega_i\right)}}\nonumber \\
&& \;\;\; = \int_{-\infty}^{\infty}{dt\,R(t)^n}.
\end{eqnarray}
For ${\cal Q} \gg 1$ the odd moments can be neglected, while the even moments are given by 
\begin{equation} \label{eq:R2m}
 {\cal R}_{2k} \approx \frac{1}{2k} ( {\cal M} \omega_c)^{-2k} \frac{\cal Q}{\omega_c}, \;\; k \ll {\cal Q} .
\end{equation}

\section{Evaluation in terms of the scattering matrix} \label{sc:constantprofile}

The Levitov-Lesovik formula (\ref{Fxicurr}) for the charge transfer statistics can
be evaluated in terms of the scattering matrix of the conductor
\cite{Lev93,Muz94,Bee01}, without explicit knowledge of the scattering states. This is possible because the current operator depends
only on the asymptotic form of the scattering states, far from the scattering
region. The formula (\ref{eq:Fxisimple}) for the momentum transfer statistics can be
evaluated in a similar way, but only if the mode profile $\bf{u}({\bf r})$ is
approximately constant over the scattering region.

\begin{figure}
\centering\epsfig{file=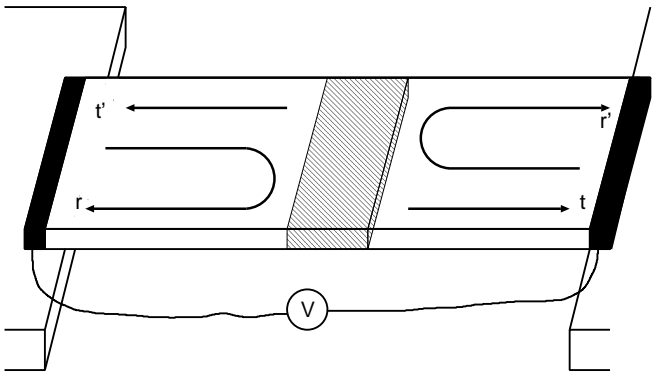,angle=0,clip=,width= \linewidth}
\caption{Sketch of a freely suspended  wire. The  matrices $t$, $t'$ and $r$, $r'$ describe transmission and reflection by a localized scatterer (shaded). A voltage $V$ drives a current through the conductor, exciting a vibration. }
\label{fig1}
\end{figure}

To this end, we first write the force operator (\ref{eq:force}) in second quantized
form using a basis of scattering states $\psi_{n,\varepsilon}({\bf r})$,
\begin{eqnarray}
&&F(t)=\int\!\!\int \frac{d\varepsilon d\varepsilon'}{2\pi}
\sum_{n,n'}e^{i(\varepsilon-\varepsilon')t}
c^{\dagger}_{n}(\varepsilon)M_{nn'}(\varepsilon,\varepsilon')
c_{n'}(\varepsilon'), \nonumber \label{FMrelation}\\
\\
&&M_{nn'}(\varepsilon,\varepsilon')=\frac{1}{m}\int d{\bf
r}\,\psi_{n,\varepsilon}^{\ast}\bigl(p_{\alpha}
u_{\alpha\beta}p_{\beta}\nonumber\\
&&\hspace{3cm}\mbox{}+[[u_{\alpha\beta},p_{\alpha}],p_{\beta}]\bigr)
\psi_{n',\varepsilon'}^{\vphantom{\ast}}.\label{Mdef}
\end{eqnarray}
The operator $c_{n}(\varepsilon)$ annihilates an electron in the $n$-th
scattering channel at energy $\varepsilon$. The mode index $n$ runs from $1$ to
$N$ (or from $N+1$ to $2N$) for waves incident from the left (or from the
right). (See Fig.\ \ref{fig1} for a diagram of the geometry and see Ref.\
\cite{But90} for the analogous representation of the current operator.)
 The commutator $[[u_{\alpha\beta},p_{\alpha}],p_{\beta}]$ can be neglected if ${\bf u}$ is smooth on the scale of the wavelength (hence if $\lambda_F/L \ll 1$).
 
We assume that the derivative $u_{\alpha\beta}$ of the mode profile vanishes in
the scattering region, so that for the scattering states we may use the
asymptotic form
\begin{equation}
\psi_{n,\varepsilon}({\bf r})=\phi_{n,\varepsilon}^{\rm in}({\bf
r})+\sum_{m}S_{mn}(\varepsilon)\phi_{m,\varepsilon}^{\rm out}({\bf
r}),\label{psiSrelation}
\end{equation}
in terms of incident and outgoing waves $\phi_{n,\varepsilon}^{\rm in,out}$
(normalized to unit current) and the scattering matrix $S_{mn}(\varepsilon)$.
Since we are neglecting the Lorentz force we may assume that $\phi_{m,\varepsilon}^{\rm out}= \phi_{m,\varepsilon}^{\rm in\ast}$. The scattering matrix has the block  structure
\begin{equation} \label{eq:S}
S=\left(\begin{array}{cc}
r&t'\\t&r'
\end{array}\right) ,
\end{equation}
with $N \times N$ transmission and reflection matrices  $t,t',r,r'$. These matrices are related by unitarity ($S=S^{\dagger}$) and possibly also by time-reversal symmetry   ($S=S^{T}$).

The operator $p_{\alpha}u_{\alpha\beta}p_{\beta}$ will couple only weakly the
incident to the outgoing waves, provided ${\bf u}$ is smooth on the scale of  $\lambda_F$, and we neglect this coupling. The matrix $M$ then separates into an
incident and outgoing part,
\begin{equation}
M(\varepsilon,\varepsilon')=M^{\rm
in}(\varepsilon,\varepsilon')+S^{\dagger}(\varepsilon)M^{\rm
out}(\varepsilon,\varepsilon')S(\varepsilon').\label{Minout}
\end{equation}
The  matrices $M^{\rm in}$ and $M^{\rm out}$ are defined as in Eq.\
(\ref{Mdef}) with $\psi$ replaced by $\phi^{\rm in}$ and $\phi^{\rm out}$,
respectively. (They are Hermitian and related by $ M^{\rm out} = M^{\rm in\ast}$.) These two matrices vary with energy on the scale of the Fermi
energy $E_{F}$, while the scattering matrix $S$ has a much stronger energy
dependence (on the scale of the Thouless energy). We may therefore replace
$M^{\rm in}$, $M^{\rm out}$ by their value at $\varepsilon=\varepsilon'=E_{F}$
and assume that the energy dependence of $M$ is given entirely by the
scattering matrix.

The force operator can similarly be separated into $F=F^{\rm in}+F^{\rm out}$,
where $F^{\rm in}$ and $F^{\rm out}$ are defined as in Eq.\ (\ref{FMrelation})
with the matrix $M$ replaced by $M^{\rm in}$ and $S^{\dagger}M^{\rm out}S$,
respectively. We now proceed in the same way as in Ref.\ \cite{Bee01} for the
current operator, by noting that the analyticity of $S(\varepsilon)$ in the
upper half of the complex plane implies simple commutation relations:
\begin{eqnarray}
&&[F^{\rm in}(t),F^{\rm in}(t')]= 0,\;\;[F^{\rm out}(t),F^{\rm
out}(t')] = 0,\;\;\forall\; t,t',\nonumber\\
&&[F^{\rm in}(t),F^{\rm out}(t')]=0\;\;{\rm if}\;t>t'.\label{Fcomm}
\end{eqnarray}
It follows that the Keldysh time ordering ${\cal T}_{\pm}$ of the force
operators is the same as the so-called input-output ordering, defined by moving
the operators $F_{\rm in}(t_{-})$ to the left and $F_{\rm in}(t_{+})$ to the
right of all other operators --- irrespective of the value of the time
arguments. The reason for preferring input-output ordering over time ordering
is that Fourier transformation from time to energy commutes with the former
ordering but not with the latter.

In the limit $t\rightarrow\infty$ different energies become uncoupled, and the
cumulant generating function takes the simple form
\begin{equation}
\ln {\cal F}(\xi)=\frac{t}{2\pi}\int d\varepsilon\,\ln\langle e^{F^{\rm
in}(\varepsilon)\xi/2}e^{F^{\rm out}(\varepsilon)\xi}e^{F^{\rm
in}(\varepsilon)\xi/2}\rangle,\label{lnFxi}
\end{equation}
entirely analogous to the input-output ordered formula for charge transfer
\cite{Bee01}. The Fourier transformed force is defined as
\begin{subequations}
\label{Finoutdef}
\begin{eqnarray}
&&F^{\rm in}(\varepsilon)=c^{\dagger}(\varepsilon)M^{\rm
in}(\varepsilon,\varepsilon)c(\varepsilon),\label{Findef}\\
&&F^{\rm
out}(\varepsilon)=c^{\dagger}(\varepsilon)S^{\dagger}(\varepsilon)M^{\rm
out}(\varepsilon,\varepsilon)S(\varepsilon)c(\varepsilon).\label{Foutdef}
\end{eqnarray}
\end{subequations}
(The operators $c_{n}$ have been collected in a vector $c$.)

The matrices $M^{\rm in,out}$ are block diagonal, 
\begin{equation} \label{eq:Minout}
M^{\rm in}=M^{\rm out\ast} =\left(\begin{array}{cc}
M_{\rm L}&0\\0&M_{\rm R}
\end{array}\right),
\end{equation}
but the $N \times N$ matrices $M_{\rm L,R}$ are in general not diagonal themselves. 
They take a simple form for a longitudinal phonon
mode, when ${\bf u}$ is  a function of  $x$ in the $x$-direction (along the
conductor), so that $u_{\alpha\beta}({\bf r})=\delta_{\alpha x}\delta_{\beta
x}u'(x)$. The commutator $[[u',p_{x}],p_{x}]$ does not contribute because
$\phi_{n}^{\rm in,out}$ is an eigenstate of $p_{x}$ (with eigenvalue
$p_{n}^{\rm in}=-p_{n}^{\rm out}\equiv p_{n})$. Hence for a longitudinal
vibration one has
\begin{subequations} \label{eq:Mlong}
\begin{eqnarray} 
&&(M_{\rm L})_{nn'}=\delta_{nn'}|p_{n}| (u_{0}-u_{\rm L}),
 \\
&&(M_{\rm R})_{nn'}=\delta_{nn'}|p_{n}| (u_{\rm R}-u_0).
\end{eqnarray}
\end{subequations}%
The value of $u(x)$ in the scattering region is denoted by $u_{0}$, while
$u_{\rm L},u_{\rm R}$ denote the values at the left and right end of the
conductor. The more complex situation of  a transverse phonon mode, when  the matrices $M_{\rm L,R}$ are no
longer diagonal, is treated in 
Ref.\  \cite{Taj02}.

We are now ready to calculate the expectation value in Eq.\ (\ref{lnFxi}). We
assume that the incident waves originate from reservoirs in thermal equilibrium
at temperature $T$, with a voltage difference $V$ between the left and right
reservoir. The Fermi function in the left (right) reservoir is $f_{\rm L}$
($f_{\rm R})$.  We collect the Fermi functions in a diagonal matrix $f$ and write
\begin{equation}
\langle
c_{n}^{\dagger}(\varepsilon)c_{n'}^{\vphantom{\dagger}}(\varepsilon')\rangle=
f_{nn'}(\varepsilon)\delta(\varepsilon-\varepsilon'),\label{ccaverage} \;\; 
f= \left(\begin{array}{cc}
f_{\rm L}&0\\0&f_{\rm R}
\end{array}\right).
\end{equation}
 All other expectation
values of $c,c^{\dagger}$ vanish. We evaluate Eq.\ (\ref{lnFxi}) with help of
the determinantal identity
\begin{equation}
\left\langle
\prod_{i}\exp(c^{\dagger}A_{i}c)\right\rangle=
\Big|\Big|1-f+f\prod_{i}e^{A_{i}}\Big|\Big|,\label{cAc}
\end{equation}
valid for an arbitrary set of matrices $A_{i}$, and the identity
\begin{equation}
\exp(S^{\dagger}AS)=S^{\dagger}e^{A}S,\label{SAS}
\end{equation}
valid for unitary $S$. The result is
\begin{equation}
\ln {\cal F}(\xi)=\frac{t}{2\pi}\int d\varepsilon\,\ln||1-f+fe^{\xi M^{\rm
in}}S^{\dagger}(\varepsilon)e^{\xi M^{\rm
out}}S(\varepsilon)||,\label{lnFxiresult}
\end{equation}
where we have also used that the two matrices $M^{\rm in}$ and $f$ commute.

At zero temperature $f_{\rm L}= \theta(E_F+e V - \varepsilon)$, $f_{\rm R}= \theta(E_F - \varepsilon)$. The energy range $\varepsilon<E_F$, where  $f_{\rm L}=f_{\rm R} =1$, contributes only to the first moment, while the energy range $E_F < \varepsilon < E_F + e V$, where $f_{\rm L} = 1$ and $f_{\rm R}=0$, contributes to all moments. For small voltages we may neglect the energy dependence of $S(\varepsilon)$ in that range. Using the block structure (\ref{eq:S}), (\ref{eq:Minout}),  of $S$,  $M^{\rm in,out}$ the generating function for the second and higher cumulants takes the form 
\begin{equation} \label{eq:momstatzero}
\ln {\cal F}(\xi)=\frac{e V t}{2\pi} \ln||r^{\dagger} e^{\xi M_{\rm
L}^*}r + t^{\dagger} e^{\xi M_{\rm
R}^*} t || + {\cal O}(\xi).
\end{equation} 
(By ${\cal O} (\xi)$ we mean  terms linear in $\xi$.) 
This determinant  can not be simplified
further without knowledge of $S$. That is a major complication relative to the
analogous formula for the charge transfer statistics \cite{Lev93}, which can
be cast entirely in terms of the transmission eigenvalues $\Gamma_n$ (eigenvalues of $t t^{\dagger}$):
\begin{eqnarray} \label{eq:chargestat}
&&\ln {\cal F}_{\rm charge}(\xi)=\frac{t}{2\pi} \int{d\varepsilon }\sum_n \ln\Big[ 1 \nonumber \\
&& \;\;\;\;\;\;\;\;\;\;\;\;\;\;\;\;\;\; \mbox{}+ \Gamma_n (e^{ e \xi}-1) f_{\rm L}(1- f_{\rm R}) \nonumber \\
&& \;\;\;\;\;\;\;\;\;\;\;\;\;\;\;\;\;\; \left. \mbox{}  + \Gamma_n (e^{ -e \xi}-1) f_{\rm R}(1- f_{\rm L})  \right] .
\end{eqnarray}
 In the case of
momentum transfer, eigenvalues and eigenvectors both play a role.

\section{Application to a one-dimensional conductor} \label{sc:single}

\subsection{Straight wire}

Further simplification of Eqs.\ (\ref{lnFxiresult}) and (\ref{eq:momstatzero}) is possible if the conductor
is so narrow that it supports only a single propagating mode to the  left and right
of the scattering region ($N=1$). 
 The scattering matrix 
then consists of scalar transmission and reflection coefficients $t,t',r,r'$
(related to each other by unitarity). 
 We consider the case of a longitudinal
vibration with
\begin{equation}
M^{\rm in}=M^{\rm out}=p_{F}\left(\begin{array}{cc}
u_{0}-u_{\rm L}&0\\0&u_{\rm R}-u_{0}
\end{array}\right),\label{M1ddef}
\end{equation}
cf.\ Eq.\ (\ref{eq:Mlong}). Because of unitarity the result depends only on the
transmission probability $\Gamma=|t|^{2}=|t'|^{2}=1-|r|^{2}=1-|r'|^{2}$,
\begin{eqnarray} \label{eq:statistics}
&&\ln {\cal F}(\xi)=\frac{t}{2\pi}\int d\varepsilon\,\ln\bigl[1 +(e^{2 \xi p_{F}(u_{\rm R}-u_{\rm L})}-1)f_{\rm L}f_{\rm R}  \nonumber \\
&&\mbox{}+\Gamma(e^{\xi p_{F}(u_{R}-u_{\rm L})}-1) [f_{\rm L}(1-f_{\rm R}) +f_{\rm R}(1-f_{\rm L})] \nonumber\\
&&\mbox{} + (1-\Gamma) (e^{2\xi p_{F}(u_{0}-u_{\rm L})}-1)f_{\rm L}(1-f_{\rm R})
\nonumber\\
&&\mbox{} +(1-\Gamma) (e^{2\xi p_{F}(u_{\rm R}-u_{0})}-1)f_{\rm R}(1-f_{\rm L}) \bigr] .\label{lnFxiresult1d}
\end{eqnarray}
At zero temperature this simplifies further to 
\begin{equation} \label{eq:Gammazero}
\ln {\cal F}(\xi) = \frac{e V t}{2 \pi} \ln[1 + \Gamma  e^{ \xi p_F (u_{\rm R}+u_{\rm L}-2 u_0)}-\Gamma ] + {\cal O}(\xi).
\end{equation}

The zero temperature statistics (\ref{eq:Gammazero}) is binomial, just as for the charge. [The generating function ${\cal F}_{\rm charge}(\xi)$ at $T=0$ is obtained from Eq.\  (\ref{eq:Gammazero})  after substitution of $p_F(u_{\rm R}+u_{\rm L}-2 u_0)$ by $e$, cf.\ Eq.\ (\ref{eq:chargestat}).] At finite temperatures one has the multinomial statistics (\ref{lnFxiresult1d}), made up of  stochastically independent elementary processes with more than two possible outcomes. The elementary processes may be  characterized by the numbers  $(n^{\rm L}_{\rm in},n^{\rm R}_{\rm in})\in \{0,1\}$ of electrons  incident on the scatterer from the left, right and the numbers   $(n^{\rm L}_{\rm out},n^{\rm R}_{\rm out})\in \{0,1\}$  of outgoing electrons to the left, right. The  non-vanishing probabilities $P[(n^L_{\rm in},n^R_{\rm in})\rightarrow (n^L_{\rm out},n^R_{\rm out})]$ of scattering events evaluate to:
\begin{eqnarray} \label{eq:table1}
P[(0,0)\rightarrow (0,0)]& =& (1-f_{\rm L})(1-f_{\rm R}), \nonumber \\
P[(0,1)\rightarrow (0,1)]& =& (1-f_{\rm L})f_{\rm R}(1-\Gamma), \nonumber \\
P[(0,1)\rightarrow (1,0)]& =& (1-f_{\rm L})f_{\rm R} \Gamma, \nonumber \\
P[(1,0)\rightarrow (1,0)]& =& f_{\rm L}(1-f_{\rm R})(1-\Gamma), \nonumber \\
P[(1,0)\rightarrow (0,1)]& =& f_{\rm L}(1-f_{\rm R})\Gamma, \nonumber \\
P[(1,1)\rightarrow (1,1)]& =& f_{\rm L} f_{\rm R}. 
\end{eqnarray}
These probabilities appear in the generating function (\ref{eq:Gammazero}), multiplied by exponentials of $\xi$ times the amount of transferred momentum.

A longitudinal vibration of a straight wire clamped at both ends would correspond to $u_{\rm L}= u_{\rm R}=0$, $u_0 \neq 0$. In that special case Eq.\ (\ref{lnFxiresult1d}) is equivalent to Eq.\ ({\ref{eq:chargestat}) for ${\cal F}_{\rm charge} (\xi)$ under the substitution $\Gamma \rightarrow 1 -\Gamma$, $2 p_F u_0 \rightarrow e$.
 In this case the multinomial statistics becomes a double-Poissonian in the limit $\Gamma \to 0$, corresponding to two independent Poisson processes originating from the left and right reservoirs \cite{Lev01}.  A longitudinal vibration is difficult to observe, in contrast to  a transverse vibration  which can be observed optically \cite{Mey88,Tre96} or magnetomotively \cite{Rou00}. However, the direct excitation of a transverse mode is not possible in a single-channel conductor, while in a multi-channel conductor (width $W$) it is smaller than the excitation of a longitudinal mode by a factor $(W/L)^2$ \cite{Taj02}.  So it would be  desirable to find a  way of coupling longitudinal electron motion to transverse vibration modes. In the following subsection we discuss how this can be  achieved  by bending the wire. 

\subsection{Bent wire} 

The bending of the wire is described as explained in Ref.\ \cite{Lan}, by means of a vector $\boldsymbol{\Omega}(s)$ that rotates the local coordinate system ${\bf e}_x(s), {\bf e}_y(s), {\bf e}_z(s)$ as one moves an infinitesimal distance $ds$ along the wire: $\delta {\bf e}_{\alpha} = \boldsymbol{\Omega} \times {\bf e}_{\alpha} \delta s$. The local coordinate $x$ is along the wire and $y,z$ are perpendicular to it. The component $\Omega_{||}$ of $\boldsymbol{\Omega}$ along the wire describes a torsion (with $|\Omega_{||}|$ the torsion angle per unit length), while the perpendicular  component $\Omega_{\bot}$ describes the bending (with $|\Omega_{\bot}|^{-1}$ the radius of curvature). 

The momentum operators and wavefunctions, written in local coordinates, depend on the bending by terms of order $\lambda_F |\boldsymbol{\Omega}|$, which we assume to be $\ll 1$.  These quantities may therefore be evaluated for a straight wire ($\boldsymbol{\Omega}=0$). The dependence on the bending of the strain tensor is of order $L  |\boldsymbol{\Omega}|$  and can not be neglected. For the interaction Hamiltonian  (\ref{eq:twoeq}) we need $\boldsymbol{\nabla} {\bf u}$ in the global coordinate system. It is obtained by differentiating the local coordinates of ${\bf u}$ as well as the local basis vectors.  A bent wire can then  be represented by a straight wire with an effective displacement ${\bf u}_{\rm eff}$ related to ${\bf u}$ (in local coordinates) by
\begin{subequations}
\begin{eqnarray} \label{eq:gradu}
 \frac{\partial}{\partial x} {\bf u}_{{\rm eff} }& =&   \frac{\partial}{\partial x} {\bf u} + \boldsymbol{\Omega}\times {\bf u}, \label{eq:gradua}  \\
\frac{\partial}{\partial y} {\bf u}_{{\rm eff} }& =&   \frac{\partial}{\partial y} {\bf u}, \;\;\frac{\partial}{\partial z} {\bf u}_{{\rm eff} } =   \frac{\partial}{\partial z} {\bf u}.
\end{eqnarray}
\end{subequations}
 The second term on the right-hand-side  of  Eq.\ (\ref{eq:gradua})  accounts for the  centrifugal force exerted by an electron moving along the bent wire.  It rotates  a transverse mode, with ${\bf u}$ pointing in radial direction, into  a fictitious longitudinal mode with $u_{{\rm eff},x} $ of order $L | \Omega_{\perp}|$. 

\begin{figure}
\centering\epsfig{file=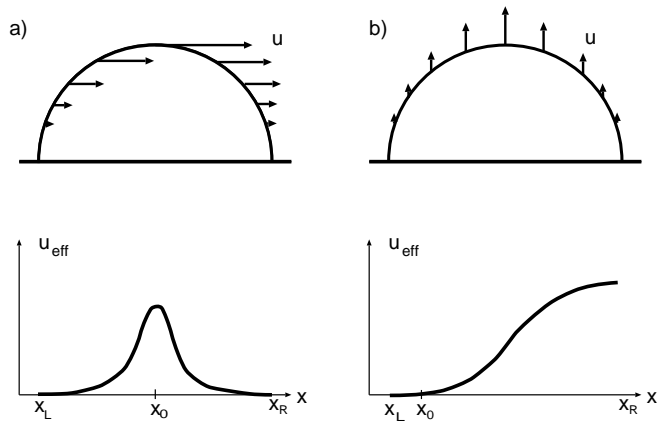,clip=,angle=0,width= \linewidth}
\caption{ Two vibration modes in a bent wire (top) and the corresponding longitudinal displacements ${\bf u}_{\rm eff}$ in the straight wire (bottom). }
\label{fig2}
\end{figure}

Fig.\ \ref{fig2} shows two vibration modes in  a bent wire  with the corresponding longitudinal component $u_{{\rm eff},x}$ of the effective displacement.  To apply the formulas of the previous subsection we need $u_{\rm L}= u_{{\rm eff},x}(x_{\rm L})$,  $u_{\rm R}= u_{{\rm eff},x }(x_{\rm R})$, and  $u_{0}= u_{{\rm eff}, x}(x_{ 0})$. The first mode, Fig.\ {\ref{fig2}a, has $u_{\rm L}=u_{\rm R}=0$ and $u_0\neq 0$. It  measures the amount of electron momentum that has been transferred to the scatterer (located at $x_0$). The statistics of this process is equivalent to the charge transfer statistics (\ref{eq:chargestat}),  as mentioned at the end of the previous subsection.

 The second mode, Fig.\ \ref{fig2}b, has  $u_{\rm L}=0$, $u_{\rm R}\neq 0$ and $u_0 \ll u_{\rm R}$ (assuming that the scatterer is located much closer to the left reservoir than to the right reservoir). It measures the amount of momentum transferred from the left to the right reservoir. Its statistics reads 
\begin{eqnarray} \label{eq:momstat}
&&\ln {\cal F}(\xi)=  \frac{t}{2 \pi} \int{d\varepsilon \ln\left\{ 1 \right.}\nonumber \\
&&\mbox{}+ (e^{2 \xi p_F u_{\rm R}}-1)[f_{\rm R}-\Gamma f_{\rm R} (1-f_{\rm L})]\nonumber\\
&&\left. \mbox{} +\Gamma(e^{ \xi  p_F u_{\rm R}} -1) [ f_{\rm L}(1-f_{\rm R}) + f_{\rm R}(1-f_{\rm L})] \right\} .\;\;\;\;\;\;
\end{eqnarray}
It cannot be reduced to the charge transfer statistics  (\ref{eq:chargestat}) by a substitution of variables, and in particular does not reduce to a double Poissonian in the limit $\Gamma \to 0$. (It remains multinomial in this limit.) 
Comparing the second cumulant $C^{(2)} $ of momentum  with the second cumulant $C^{(2)}_{\rm charge}$ of charge [the terms of order $\xi^2$ in Eqs.\ (\ref{eq:chargestat}) and (\ref{eq:momstat})], we find (setting $u_{\rm R} \equiv 1$)
\begin{equation} \label{eq:statdiff}
 C^{(2)} - (p_F/e)^2 C^{(2)}_{\rm charge}  = \frac{2}{\pi} t p_F^2 k_B T (1-\Gamma).
\end{equation}
The difference vanishes at zero temperature, in accordance with Eq.\ (\ref{eq:Gammazero}).
It is independent of the voltage (as long as the energy dependence of $\Gamma$ can be ignored), so the difference is an equilibrium property.

 Eq.\ (\ref{eq:statdiff}) can be given a physical interpretation by grouping the electrons  to the right  of the scattering region into $n_>$ right movers and $n_<$ left movers.  The momentum transfer to the right reservoir is proportional to  the sum $n_>+n_<$ while the charge transfer is proportional to the difference $n_>-n_<$, hence 
\begin{eqnarray}
 && C^{(2)}  - \frac{p_F^2}{e^2} C^{(2)}_{\rm charge} \propto  \langle \langle (n_>+n_<)^2 \rangle \rangle - \langle \langle (n_>-n_<)^2 \rangle \rangle \nonumber \\
&& =\mbox{}  4(\langle n_> n_< \rangle - \langle n_> \rangle \langle n_< \rangle).
\end{eqnarray}
We see that the difference measures  correlations between left and right-moving electrons. Such correlations are due to electrons that are backscattered with probability $1-\Gamma$. Eq.\ (\ref{eq:statdiff}) describes  the variance in the number of such backscattered electrons, given that electrons in an energy range $k_B T$ leave the right reservoir independently of each other.

\section{Evaluation in terms of the Keldysh Green function} \label{sc:Keldysh}

A scattering approach as in Sec.\ \ref{sc:constantprofile} is not possible if the displacement ${\bf u}({\bf r})$  varies in the scattering region. 
Time ordering  then no longer reduces to input-output ordering and we need the Keldysh technique to make progress \cite{Rammer}. Following the analogous formulation of the charge counting statistics \cite{Naz99}, we write the generating function  (\ref{eq:Fxisimple}) as a 
 single exponential of an integral  along the Keldysh time contour, 
\begin{subequations}
\begin{eqnarray} \label{eq:Sel}
 {\cal F}(\xi) =  \left\langle {\cal T}_{\pm} \exp\left[{\textstyle{\frac{1}{2}}} \xi   \int_0^t{dt'{ \int d{\bf r}\,{ F_{\pm}({\bf r},t')} }}\right] \right\rangle,  \\
F_{\pm}({\bf r},t)= \frac{1}{m} \sum_{\sigma = \pm}  \psi_{\sigma}^{\dagger}({\bf r},t) p_{\alpha}u_{\alpha\beta}({\bf
r})p_{\beta}\psi_{\sigma}({\bf r},t) . 
\end{eqnarray}
\end{subequations}%
We have written the force operator in second quantized form, as in Eq.\ (\ref{FMrelation}), but do not assume that the electron field operator $\psi_{\pm}({\bf r},t) \equiv \psi({\bf r},t_{\pm}) $ takes its asymptotic form in terms of incident and outgoing states.

The generating function can be expressed in terms of the Keldysh Green function $G$, 
\begin{eqnarray}  \label{eq:partialSel}
\frac{d}{d \xi} \ln {\cal F}(\xi)& = &
 \frac{i}{2m} \sum_{\sigma=\pm} \sigma\int_0^t{dt'\int{d \boldsymbol{R}\, u_{\alpha \beta}(\boldsymbol{R}) }}\nonumber \\
&& \!\!\!\!\!\!\!\! \mbox{}\times  \frac{\partial^2}{\partial { r}_{\alpha} \partial { r}_{\beta}} G_{\sigma\sigma}({\bf R},{\bf r},t',t';\xi)\Big|_{{\bf r}=0}.
\end{eqnarray}
The Green function $G_{\sigma\sigma'} $ is a $2 \times 2$ matrix in the indices $ \sigma,\sigma' \in \{+,-\}$ that assure the correct time ordering of the operators. It is defined by 
\begin{widetext}
\begin{equation} \label{eq:Gxi} 
{G}_{\sigma\sigma'}({\bf R},{\bf r},t,t';\xi) = \frac{-i \sigma  \left\langle {\cal T}_{\pm}  {\psi}_{{\sigma} }({\bf R}+{\textstyle{\frac{1}{2}}} {\bf r},t) {\psi}^{ \dagger}_{{\sigma'} }({\bf R}-{\textstyle{\frac{1}{2}}} {\bf r},t')  \exp\left[ {\textstyle{\frac{1}{2}}} \xi \int_0^t{dt'  \int{d{\bf r'}\; F_{\pm}({\bf r'},t')   }} \right]   \right\rangle }{ \left\langle  {\cal T}_{\pm}\exp\left[ {\textstyle{\frac{1}{2}}} \xi \int_0^t{dt'  \int{d{\bf r'}\; F_{\pm}({\bf r'},t')   }} \right]   \right\rangle}.
\end{equation} 
\end{widetext}

\section{Application to a diffusive conductor}
 \label{sc:diffusive}
 
We apply the formalism of Sec.\ \ref{sc:Keldysh} to  the example of  diffusive electron transport through a freely suspended disordered wire. The semi-classical calculation  of the transverse momentum noise in this geometry was done in Ref.\ \cite{Shy02}, so we can compare results.

 For long detection times we may assume that the Green function (\ref{eq:Gxi}) depends only on the difference $\tau=t-t'$ of the time arguments. A Fourier transform gives
\begin{equation}
{G}_{\sigma \sigma'}(\boldsymbol{R} ,\boldsymbol{p},\varepsilon;\xi) = \int{d \boldsymbol{r}\int{d \tau \; e^{-i \boldsymbol{p}\cdot \boldsymbol{r}- i \varepsilon \tau} {G}_{\sigma \sigma'}(\boldsymbol{R},\boldsymbol{r},\tau;\xi)}}.
\end{equation}
We write ${\bf p}=|{\bf p}| {\bf n}$ and  use  the fact that in the semi-classical limit the Green function is peaked as a function of the absolute value $|{\bf p}|$ of the momentum. Integration over this variable yields the  semi-classical Green function \cite{Rammer}
\begin{equation}
G_{\sigma \sigma'}(\boldsymbol{R},{\bf n},\varepsilon;\xi) = \frac{i}{\pi} \int{d\varepsilon' \; {G}_{\sigma\sigma'}(\boldsymbol{R},\boldsymbol{n} \sqrt{2m \varepsilon'},\varepsilon;\xi)}.
\end{equation}
We next  make the diffusion approximation, expanding the ${\bf n}$-dependence in spherical harmonics,
\begin{eqnarray} \label{eq:diffexp}
&& G_{{\sigma \sigma'}} (\boldsymbol{R},{\bf n},\varepsilon;\xi) = G^{(0)}_{\sigma \sigma'}(\boldsymbol{R},\varepsilon;\xi) + {n}_{\alpha} G_{\alpha \sigma \sigma'}^{(1)}(\boldsymbol{R},\varepsilon;\xi) \nonumber \\
&&\;\;\; \;\;\;\; \mbox{}+ (n_{\alpha} n_{\beta} - {\textstyle{\frac{1}{3}}} \delta_{\alpha \beta}) G^{(2)}_{\alpha \beta \sigma \sigma'}(\boldsymbol{R},\varepsilon;\xi).
\end{eqnarray}
Substituting Eq.\ (\ref{eq:diffexp})   into Eq.\ (\ref{eq:partialSel}) we find 
\begin{eqnarray} \label{eq:partialsigma}
&&\frac{d}{d \xi} \ln F(\xi) = {{\textstyle{\frac{1}{2}}}}  t  E_F \nu \sum_{\sigma=\pm} {\sigma} \int{d\varepsilon \int{d\boldsymbol{R} \;  u_{\alpha \beta}(\boldsymbol{R})}}\nonumber \\
&& \mbox{} \times   \left[\frac{1}{3} \delta_{\alpha \beta}G^{(0)}_{{\sigma \sigma}}(\boldsymbol{R},\varepsilon;\xi)  + \frac{2}{15} G^{(2)}_{\alpha \beta\sigma \sigma}(\boldsymbol{R},\varepsilon;\xi) \right],
\end{eqnarray}
where $\nu=p_F^2/2\pi^2 v_F$ is the density of states. 

The equation of motion for the semi-classical  Green function in the diffusion   approximation is derived in the same way as for the charge statistics \cite{Naz99}. We find
\begin{equation} \label{eq:eqmotion}
 2 l {n}_{\alpha} \frac{\partial}{\partial R_{\alpha}}  G + 
 [G^{(0)}, G] +  \xi  p_F l   u_{\alpha\beta} n_{\alpha} n_{\beta}  [\tau_3, G]=0.
\end{equation}
The length $l$ is the mean free path, assuming isotropic impurity scattering. The commutators $[..,..]$ are taken with respect to the Keldysh indices $\sigma, \sigma'$ and $\tau_3$ is the third Pauli matrix in these indices.  The Green function satisfies the normalization condition  $G^2=1$  that is respected by the differential equation (\ref{eq:eqmotion}). The  boundary conditions  at the left and right  ends of the wire  are \cite{Naz99}
\begin{eqnarray} \label{eq:xiG}
G_{\rm L}=  \left( \begin{array}{cc} 1 - 2 f_{\rm L} &   2 f_{\rm L} \\  2 - 2f_{\rm L} &  2 f_{\rm L}-1 \end{array} \right), \nonumber \\
G_{\rm R}=  \left( \begin{array}{cc} 1 - 2 f_{\rm R} &   2 f_{\rm R} \\  2 - 2f_{\rm R} &  2 f_{\rm R}-1 \end{array} \right).
\end{eqnarray}

By projecting Eq.\ (\ref{eq:eqmotion}) onto spherical harmonics we find that, to leading order in $l/L$,  the second harmonic $G^{(2)}$ depends only on the zeroth harmonic $G^{(0)}$:
\begin{equation} \label{eq:sigma}
 G^{(2)}_{\alpha \beta} =\frac{\xi}{2}  p_F l ( u_{\alpha\beta} - \frac{1}{3} \delta_{\alpha \beta}  u_{\gamma \gamma} ) G^{(0)}[\tau_3,G^{(0)}] [1+ {\cal O}(l/L)^2].
\end{equation}
Combining this relation with Eq.\ (\ref{eq:partialsigma}) we see that the momentum statistics of a transverse mode, with $u_{xx}=0$, $u_{xy}\neq 0$,  follows from
\begin{eqnarray} \label{eq:partials}
\frac{d}{d \xi} \ln F(\xi)& =& \frac{\xi}{30} t p_F l E_F \nu   \sum_{\sigma,\alpha,\beta} \int{d\varepsilon \int{d\boldsymbol{R} \;  u^2_{\alpha \beta} }} \nonumber \\ 
&& \times \left(\tau_3 G^{(0)} [ \tau_3,G^{(0)}]\right)_{\sigma \sigma}. 
\end{eqnarray}

 It remains to compute $G^{(0)}$.  To calculate $\ln{\cal F}$ to order $\xi^2$, that is to calculate the variance $C^{(2)}$ of the force noise, it is sufficient to know  $G^{(0)}$  for $\xi=0$. The solution to the unperturbed diffusion equation (\ref{eq:eqmotion}) is  known \cite{Naz99},
\begin{equation}
 G^{(0)}({\bf R},\varepsilon;\xi=0)=  \left( \begin{array}{cc} 1 - 2 f({\bf R},\varepsilon) &   2 f({\bf R},\varepsilon) \\  2 - 2f({\bf R},\varepsilon) &  2 f({\bf R},\varepsilon)-1 \end{array} \right)
,
\end{equation}
where  $f(\boldsymbol{R},\varepsilon) = f_{\rm L}(\varepsilon) + (x/L) [f_{\rm R}(\varepsilon) -f_{\rm L}(\varepsilon)]$. (The coordinate $x$ runs along the wire, from $x=0$ to $x=L$.) We find
\begin{equation} \label{eq:transverse}
 C^{(2)} =    t \frac{16}{15} p_F l E_F \nu  A \int_0^L{d x  d\varepsilon \,u^2_{xy}(x) f(x,\varepsilon) [1-f(x,\varepsilon)]},
\end{equation}
with $A$ the cross-sectional area of the wire. 
This is the same result as in Ref.\ \cite{Shy02}.

More complicated networks of diffusive wires, including tunnel barriers or point  contacts, can be treated in the same way. In such situations  the unperturbed  Green function $G^{(0)}({\bf R},\varepsilon;\xi=0)$  can  be determined using Nazarov's circuit theory   \cite{circuit} and then substituted into  Eq.\ (\ref{eq:partials}).

\section{Conclusion} \label{sc:conclusion}

We conclude by estimating the order of magnitude of the cumulants of the displacement distribution $P(Q)$ of a vibrating current-carrying wire. For an oscillator with a large quality factor  only the even order cumulants $\langle \langle Q^{2k} \rangle \rangle$ are appreciable, given in good approximation by 
\begin{equation} \label{eq:Q2m}
\langle \langle Q^{2k} \rangle \rangle \approx \frac{1}{2k} ({\cal M} \omega_0)^{-2k} \frac{\cal Q}{\omega_0} \lim_{t \to \infty} \frac{1}{t} \langle \langle \triangle P(t)^{2k} \rangle \rangle,
\end{equation}
cf.\ Eqs.\ (\ref{eq:Qtom}) and (\ref{eq:R2m}). The cumulants of transferred momentum $\triangle P$ have been calculated for a single-channel conductor with a localized scatterer in Sec.\ \ref{sc:single}. At zero temperature one has
\begin{eqnarray} \label{eq:delP}
 && \lim_{t \to \infty} \frac{1}{t} \langle \langle \triangle P(t)^{2k} \rangle \rangle = \frac{eV}{2\pi \hbar} p_F^{2k} (u_{\rm R} + u_{\rm L} - 2 u_0)^{2k} \nonumber \\
&& \;\; \mbox{} \times \frac{d^{2k}}{d\xi^{2k}} \ln[1+\Gamma(e^{\xi}-1)]\Big|_{\xi=0},
\end{eqnarray}
cf.\ Eq.\ (\ref{eq:Gammazero}). (We have reinserted Planck's constant $\hbar$ for clarity.)

Combining Eqs.\ (\ref{eq:Q2m}) and (\ref{eq:delP}) we see that in order of magnitude $\langle \langle Q^{2k} \rangle \rangle \simeq (eV{\cal Q}/\hbar \omega_0)  (p_F/{\cal M} \omega_0)^{2k}$.  Inserting parameter values (following Ref.\ \cite{Sch95}) $V=1\,{\rm mV}$, ${\cal Q}= 10^3$, $\omega_0/2\pi=5\,{\rm GHz}$, $p_F = 5\cdot \,10^{-24} \,{\rm Ns}$, and ${\cal M} = 10^{-20}\, {\rm kg}$, we estimate
 \begin{equation}
  \langle \langle Q^{2k} \rangle \rangle^{1/2k} \approx  10^{4/2k} \times 10^{-4} \mbox{\AA}.
\end{equation}
Detectors with a $ 10^{-4} \mbox{\AA}$  sensitivity have been proposed \cite{Ble01}. 
 For a measurement of higher order cumulants one would want cumulants of different order to be  of roughly the same magnitude. This can be achieved by choosing the number  $ eV{\cal Q}/\hbar \omega_0$     not too large. For the parameters chosen above,   $  \langle \langle Q^{4} \rangle \rangle^{1/4}/ \langle \langle Q^{2} \rangle \rangle^{1/2}    \approx 0.1\,$ .                                                               

The theory presented in this work is more than  a
framework for the calculation of higher order cumulants in the
momentum transfer statistics. It also provides for a formalism 
to treat quantum effects in electromechanical noise. A first
application, to quantum size effects in a constriction, has
been realized \cite{Taj02}. Other applications,
including resonant tunneling, superconductivity, and interaction
effects, are envisaged.

\begin{acknowledgements}
This research was supported by the ``Ne\-der\-land\-se
or\-ga\-ni\-sa\-tie voor We\-ten\-schap\-pe\-lijk On\-der\-zoek'' (NWO)
and by the ``Stich\-ting voor Fun\-da\-men\-teel On\-der\-zoek der
Ma\-te\-rie'' (FOM).
\end{acknowledgements}

\appendix{
\section{Derivation of the Unitary Transformation (2.4)} \label{app:U}

We demonstrate that the operator $U$ given in  Eq.\ (\ref{Uresult}) has the desired property  (\ref{eq:UVUdagger}) of eliminating the phonon displacement from the ion potential.
 By expanding the exponential in  Eq.\ (\ref{Uresult}) we calculate the effect of $U$  on a one-electron and one-phonon  wavefunction in  the position space representation:
\begin{equation} \label{eq:Ueffect}
  {U} \psi(\boldsymbol{r},q) = \; ||J ||^{1/2}\; \psi[\boldsymbol{r} - q \boldsymbol{u}(\boldsymbol{r}),q].
\end{equation}
We  prove Eq.\  (\ref{eq:UVUdagger}) by calculating matrix elements,
 \begin{widetext}   
\begin{eqnarray}
\langle  \psi_1 | U^{\dagger} V[\boldsymbol{r} - Q \boldsymbol{u}(\boldsymbol{r})]U |\psi_2 \rangle & =& 
 \int{d {\bf r} \int{dq \,|| J|| \,\psi_1^*[{\bf r} - q {\bf u}({\bf r}),q]  V[\boldsymbol{r} - q \boldsymbol{u}(\boldsymbol{r})] \psi_2[{\bf r} - q {\bf u}({\bf r}),q]} } \nonumber \\
& = & \int{d \tilde{\bf r} \int{dq \; \psi_1^*(\tilde{\bf r} ,q)  V(\tilde{\boldsymbol{r}}) \psi_2(\tilde{\bf r} ,q) }} =   \langle \psi_1 |V| \psi_2 \rangle.
\end{eqnarray} 
\end{widetext}    
The unitarity of $U$ follows as the special case $V \equiv 1$.

We now justify the replacement of $\tilde{\rho}=U^{\dagger}\rho U$ with
$\rho$ and $\tilde{A}=U^{\dagger}AU$ by $A$ in the generating function
(\ref{FHint}), in the limit of a long detection time $t$.
Since $Q$ commutes with $U$, it is sufficient to consider $A=P$. (Then
$\tilde{A}=A(Q,\tilde{P})$ in the more general case that $A$ is a
function of both $Q$ and $P$.) To first order in the displacement one has 
\begin{equation} 
\tilde{P}=P-\Pi+{\cal O}({\bf u}^{2}). 
\end{equation} 
The difference between $\tilde{P}$ and $P$ is of the order of the total
momentum $\Pi$ inside the wire, which is $t$-independent in a stationary
state. Since the expectation value (as well as higher cumulants) of $P$
increases linearly with $t$, we can neglect the difference between
$\tilde{P}$ and $P$ for large $t$.

To justify the replacement of $\tilde{\rho}$ by $\rho$ we note that the
effect of $U$ on the initial state is to shift the electron coordinates
by the local phonon displacement [cf.\ Eq.\ (\ref{eq:Ueffect})]. This initial shift
has only a transient effect and can be neglected for large $t$.

\section{effective mass approximation} \label{app:effmass}

We start with the Hamiltonian (\ref{eq:twoeq}) with $V=V_{\rm lat}+V_{\rm imp}$. In
the absence of any deformation of the periodic lattice one has, in the
effective mass approximation, 
\begin{equation}  \label{eq:effmass}
\frac{1}{2m}{\bf
p}^{2}+V_{\rm lat}({\bf r})=\frac{1}{2m^*}{\bf p}^{\ast 2}.
\end{equation} 
The quasimomentum operator ${\bf p}^{\ast}$ is defined in
terms of the Bloch function $g({\bf r})$ by ${\bf p}^{\ast}=-ig {\boldsymbol{ \nabla}} g^{-1}$.
We seek a similar approximation to the same Hamiltonian in a distorted lattice, assuming that ${\bf u}$ is sufficiently smooth that we can neglect 
 derivatives of the shear tensor $u_{\alpha \beta}$.
The Hamiltonian (\ref{eq:twoeq}) (for one electron) then has the form
\begin{eqnarray}
H&=& \frac{1}{2m}p_{\alpha} ( \delta_{\alpha \beta} - 2 Q  u_{\alpha \beta}  )  p_{\beta} + V_{\rm lat} + V_{\rm imp}  \nonumber \\
&& \mbox{}  -\frac{1}{M} P \Pi +\Omega  a^{\dagger}a.
\end{eqnarray}
For small displacements $Q$ the real symmetric matrix $X_{\alpha \beta} = \delta_{\alpha\beta} - 2 Q u_{\alpha \beta} $ is positive definite. We can therefore factorize ${\bf X}={\bf TT}^T$, with  ${\bf T}$ real.  We change coordinates to  $\tilde{\bf r} = {\bf T}^{-1} {\bf r} $ and find
\begin{eqnarray}
H&=&  - \frac{1}{2m} \frac{\partial}{\partial \tilde{r}_{\alpha}}  \frac{\partial}{\partial \tilde{r}_{\alpha}}   + V_{\rm lat}({\bf T} \tilde{\bf r})+  V_{\rm imp}({\bf T} \tilde{\bf r})  \nonumber \\
&& \mbox{}-  \frac{1}{M} P \Pi +\Omega a^{\dagger}a.
\end{eqnarray}

We now make the  assumption of a deformation independent effective mass \cite{Kon84,Fik78}, that is to say, we assume that the Hamiltonian with the  distorted lattice potential $V_{\rm lat}({\bf T} \tilde{\bf r})$ is approximated as in  Eq.\ (\ref{eq:effmass}) with distorted Bloch functions, but the same effective mass $m^*$. 
Hence
\begin{eqnarray}
H&=& - \frac{1}{2m^*}  \left[g({\bf T} \tilde{\bf r})  \frac{\partial}{\partial \tilde{r}_{\alpha}}    \frac{1}{g({\bf T} \tilde{\bf r})} \right]^2    + V_{\rm imp}({\bf T} \tilde{\bf r})   \nonumber \\
&& \mbox{}  - \frac{1}{M} P \Pi +\Omega a^{\dagger}a.
\end{eqnarray}
Transforming back to the original coordinates we arrive at the Hamiltonian
\begin{eqnarray}
H&=&  \frac{1}{2m^*}  p^*_{\alpha} ( \delta_{\alpha \beta} - 2 Q u_{\alpha \beta}  ) p^*_{\beta}  +  V_{\rm imp}  -   \frac{1}{M} P \Pi +\Omega a^{\dagger}a \nonumber \\ 
\end{eqnarray}
given in Sec.\ \ref{sc:Formulation}.

}

\end{document}